\documentclass[conference]{IEEEtran}

\usepackage{cite}
\usepackage[cmex10]{amsmath}
\usepackage{amssymb}
\usepackage{bm}
\usepackage{graphicx} 
\DeclareGraphicsExtensions{.jpg,.eps,.gif,.png,.pdf}
\graphicspath{{img/png/} {img/eps/}{img/pdf/}{matlab/fig/}}
\usepackage{caption}
\usepackage{subcaption}
\usepackage{stfloats}
\usepackage{epstopdf}
\usepackage{url}

\usepackage{upgreek}
\usepackage{flushend}
\usepackage{color}
\usepackage{multirow}

\def\PL{\text{PL}}

\begin{document}
%
\title{Improving Smartphone Battery Life Utilizing Device-to-device Cooperative Relays \\
Underlaying LTE Networks}
\author{Tuan Ta, John S. Baras, Chenxi Zhu\\
Institute for Systems Research\\
University of Maryland, College Park, USA\\
\{tta, baras, czhu2001\}@umd.edu}

\maketitle

\begin{abstract}
The utility of smartphones has been limited to a great extent by their short battery life. In this work, we propose a new approach to prolonging smartphone battery life. We introduce the notions of \emph{valueless} and \emph{valued battery}, as being the available battery when the user does or does not have access to a power source, respectively. We propose a cooperative system where users with high battery level help carry the traffic of users with low battery level. Our scheme helps increase the amount of valued battery in the network, thus it reduces the chance of users running out of battery early. Our system can be realized in the form of a proximity service (ProSe) which utilizes a device-to-device (D2D) communication architecture underlaying LTE. We show through  simulations that our system reduces the probability of cellular users running out of battery before their target usage time (probability of outage). Our simulator source code is made available to the public.
\end{abstract}


\section{Introduction}
\label{sec:intro}

The future of smartphones surrounding us is no secret. Smartphones have been outselling PC since the end of 2010 \cite{Guardian_ol}. 
The number of smartphones as well as the traffic generated by them are increasing dramatically \cite{Cisco_ol}. However, short battery life has been a major limiting factor for the utility of smartphones as battery technology could not keep pace. A lot of research effort has been put into designing energy efficient protocols and networks to make best use of the available battery capacity. Factors contributing to power consumption in a smartphone are broken down and studied in detail \cite{Carroll10, Shye10}. It is shown that radio communications, together with the backlit screen, consume the most energy, significantly higher than other components such as processor, memory. We are interested in the problem of prolonging smartphones' battery life by reducing power consumption due to communication. Solutions to this problem have been proposed in all layers, ranging from efficient designs for applications, heterogeneous or mixed cell deployment, energy-aware scheduling, to MIMO.

All previous works attempting to prolong battery life in wireless networks have either considered a single device, or tried to minimize the total power consumption in cooperative schemes. In the context of cellular networks (LTE in particular), rather than reduction, we propose an entirely new approach: \emph{redistributing} the existing energy to increase usage time. First, we introduce the notion of \emph{valued battery}, defined as the battery of the smartphone when the user is active and does not have access to a power source. Conversely, \emph{valueless battery} is defined as the remaining battery of the smartphone after the usage period, when the user gets access to a power source. \emph{Outage events} are instances when the user runs out of (valued) battery before his target usage time. Since the usage patterns of the users vary, the value of their batteries also varies. Our system takes advantage of the wide range of battery value created by this diversity of usage. By enabling cooperation, we allow users to spend their valueless battery to save someone else's valued battery, reducing the probability of their outage events as a result.
The physical mechanism, our means for ``distributing'' battery, is device-to-device cooperative relay underlaying cellular networks.

Device-to-device (D2D) communications underlaying cellular networks are practices of creating direct links between cellular users. 
Here we consider D2D operating on licensed spectrum as proposed in 3GPP release 12 work item \cite{tr22.803}. The benefit of allowing direct device communication on license spectrum over currently available means such as Wifi or bluetooth is that D2D interference is controlled. As a result, the bandwidth and QoS of the communications can be guaranteed. Furthermore, D2D operation can be transparent to the users. Since both D2D devices already have a secure connection to the cellular network, a secure D2D connection can be set up automatically (as compared to manual pairing in Wifi and bluetooth). A survey of D2D communications underlaying cellular networks can be found in \cite{Fodor12}. 
The property of a D2D connection that is of most importance to us is that it consumes significantly less power than a cellular link. This is because on the uplink, the phone needs to cover a much shorter distance to reach a D2D neighbor than to reach a base station. 

In our system, a user with low battery requests help from his neighbors. A selected neighbor acts as a relay through a D2D link established with the requester, bearing the cost of the cellular link. Effectively, the neighbor ``lends'' the requester battery for that transaction. The randomness of user usage ensures that with high probability, the helper will run low on battery at some other time and receive help. Our system works better as the number of users increases. Thus we benefit from the growing trend of smartphone usage.

This paper is structured as follows. In Section~\ref{sec:d2d}, D2D communication underlaying cellular networks and its advantages are outlined. We describe the operations of our system and illustrate how it can be implemented as a 3GPP proximity service in Section~\ref{sec:system_description}. In section~\ref{sec:performance_evaluation} we discuss our performance evaluation framework. We show the simulation results in Section~\ref{sec:simulation}. Finally, we conclude in Section~\ref{sec:conclusion}.

\section{D2D communications underlaying cellular networks}
\label{sec:d2d}
\begin{figure}
\centering
\includegraphics[width=0.25\textwidth]{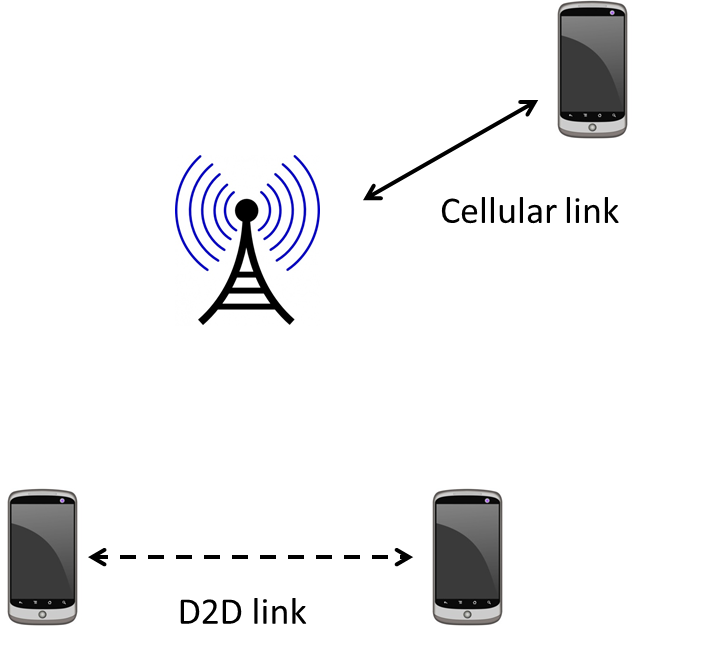}
\caption{A simple scenario with one D2D pair and one cellular user.}
\label{fig:d2d}
\end{figure}

A simple scenario with a D2D pair and a regular cellular user is illustrated in Fig~\ref{fig:d2d}. The base station (eNodeB in LTE terminology) wants to schedule these two links concurrently. It has two choices: 1) using dedicated resources for each link, and 2) allowing both links to share the same time-frequency resource. In the first case, there is no interference between the D2D link and the cellular link, therefore the D2D link has the same bandwidth and QoS as a normal cellular link. In the second case, depending on whether the uplink or downlink resource is shared, different interference management methods need to be applied. D2D interference management is an important topic which has gained a lot of attention due to the potential of allowing local traffic at no cost to regular users. Proposed solutions range from power control to MIMO precoder design \cite{Janis09_3, Hakola10, Min11, Elkotby12}. For the purpose of this work, we assume that scheduling and interference management techniques exist, and a D2D link can be established when the 2 terminals are within range.

The advantage of a D2D link compared to a cellular link is that it covers a much shorter distance. The main component of the signal energy loss over the wireless channel is the distance-related path loss. Using the nominal values in Table~\ref{tab:path_loss}, we calculate the path loss of the D2D link versus the cellular link according to both UMTS channel model \cite{tr30.03} and IST WINNER II channel model \cite{winner}. In addition, the eNodeB receiver has better gain (14 dBi) and lower noise figure (5 dB) compared to the UE receiver (0 dBi, 9 dB) \cite{tr25.814}. Table~\ref{tab:pl_results} shows that under similar fading conditions and ignoring shadowing, to get the same SNR at the receiver, the cellular UE needs to spend 3 to 4 orders of magnitude more transmission power than the D2D transmitter. 

\begin{table}
\centering
\caption{Nominal values for path loss models}
\begin{tabular}{|c|c|}
\hline
\textbf{Parameter} & \textbf{Value} \\ \hline
UE - macro eNodeB distance & 300 m\\ \hline
D2D UEs distance & 10 m\\ \hline
Carrier frequency & 2 GHz\\
\hline
\end{tabular}
\label{tab:path_loss}
\end{table}
\begin{table}
\centering
\caption{Path loss results (in $\textnormal{dB}$)}
\begin{tabular}{|c|c|c|c|c|}
\hline
\textbf{Channel model} & \textbf{Cellular} & \textbf{D2D} & \textbf{PL diff.} & \textbf{Tx power diff.} \\ \hline
UMTS& 127 & 67 & 60 & 42\\ \hline
WINNER II & 122 & 73 & 49 & 31\\ 
\hline
\end{tabular}
\label{tab:pl_results}
\end{table}

In addition, it has also been shown that in cellular networks, uplink transmission power can be an order of magnitude higher than downlink reception power \cite{Huang12}. The above analysis justifies that using D2D relay provides significant energy saving as compared to using a cellular link.

\section{System description}
\label{sec:system_description}
In this section we describe the operation of our system as well as its overhead and security implications.

\subsection{Operation}
Our system can be implemented as a proximity service (ProSe) as described in \cite{tr22.803}. We call it the Battery Deposit Service (BDS). The name is derived from the fact that when a user spends his valueless battery to save another user's valued battery, it can be thought of as ``depositing'' battery into the network. The user whose valued battery is conserved can be thought of as ``withdrawing'' battery from the network. The concepts of depositing and withdrawing are used to signify the fact that the benefit of a helper needs not be immediate or reciprocal. In other words, a user receiving help can repay, at a later time, a different user than the one who helps him. This way BDS benefits from the large population of users in the network.

In order to enable operator-controlled device and service discovery as well as D2D connection set up, the Evolved Packet Core (EPC) must include additional functionalities to manage D2D services. One method to provide those functionalities is suggested in \cite{Raghothaman13}, where two new entities are added: D2D Server and Application Server (AppSer). The D2D Server is responsible for maintaining D2D-enabled devices' identity, coordinating the establishment of D2D connections, as well as storing usage records for charging purpose. The AppSer performs application/service-specific tasks (because one UE can use multiple D2D services at the same time, with BDS being one of those services). 

\begin{figure}
	\centering
	\includegraphics[width=0.5\textwidth]{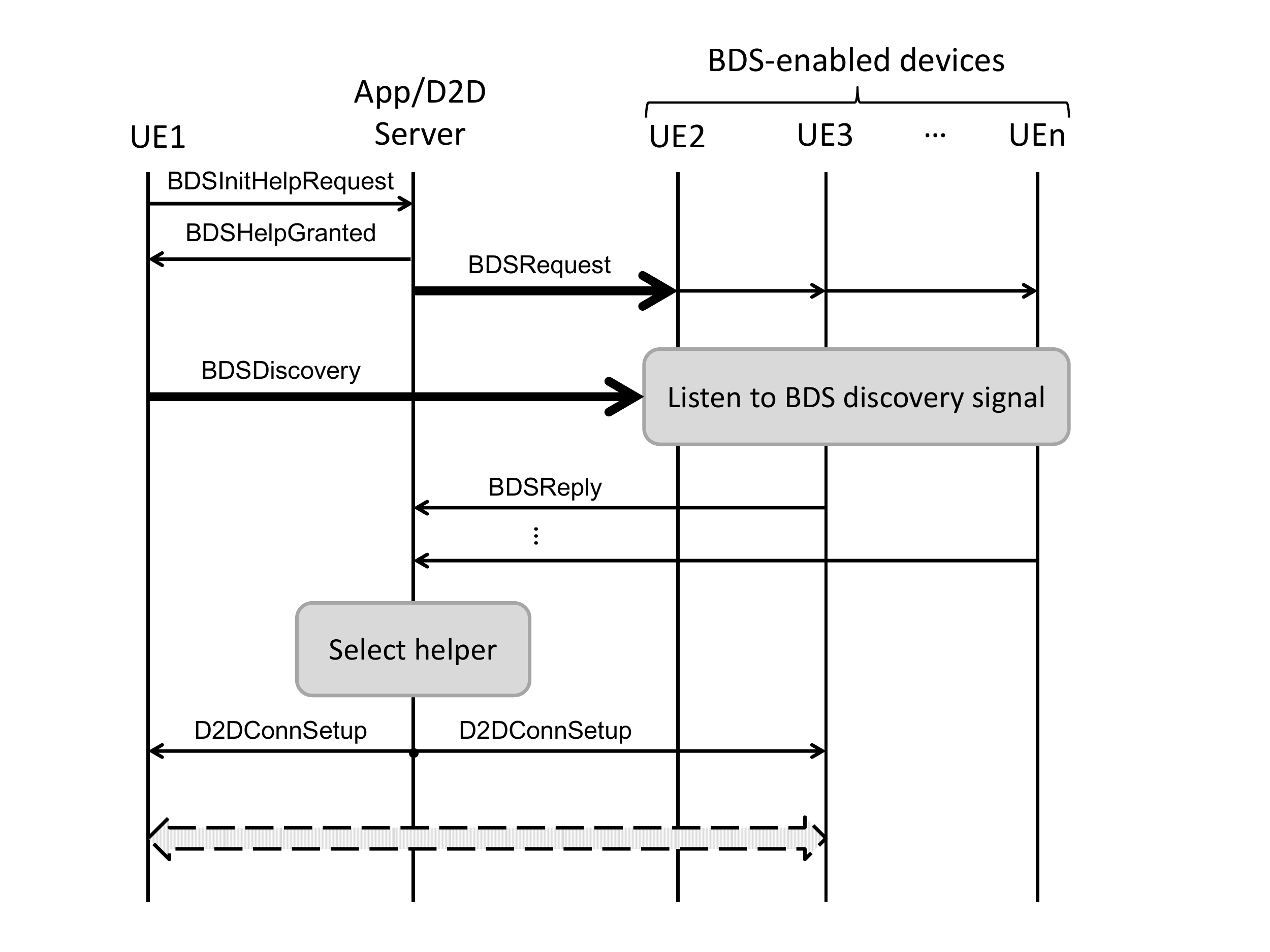}
	\caption{Battery Deposit Service operational flow. Here UE1 is the helpee and UE3 is the selected helper. UE2 is BDS-enabled but is not in the proximity of UE1.}
	\label{fig:bds_flow}
\end{figure}

As ProSe is an active working item within 3GPP and there is no standard on how a service should be defined yet, we opt to describe the operational flow of BDS instead of the exact signaling formats. This operational flow is illustrated in Figure~\ref{fig:bds_flow}. When a UE's battery level goes below a threshold $\gamma_1$ it considers looking for help. When it verifies that its channel condition is bad (downlink Reference Signal Received Power is less than a threshold), and receiving help is beneficial, the UE sends \texttt{BDSInitHelpRequest} to the AppSer. Let us name this UE1. The AppSer responds with a \texttt{BDSHelpGranted} message where a time-frequency resource is allocated to UE1 for a neighbor discovery signal. After receiving the acknowledgment from UE1 (not shown in Figure~\ref{fig:bds_flow}), the AppSer sends a multicast message \texttt{BDSRequest} to all BDS-enabled UEs in the cell. This \texttt{BDSRequest} message includes the scheduled resource for UE1's discovery signal, \texttt{BDSDiscovery}. All available helpers, whose battery level is above another threshold $\gamma_2$, listen to this resource unit. Any helper who is able to hear UE1's discovery signal will report to the AppSer through a message \texttt{BDSReply}. 

After receiving the list of potential helpers for UE1, the AppSer runs a helper selection algorithm to determine the helper for UE1, together with the duration of this association. In our framework, the helper selection algorithm can be flexibly designed to achieve different goals. To assist the selection algorithm, additional information about the potential helpers can be passed to the AppSer through \texttt{BDSReply}. Some possible selection algorithms include: 
\begin{itemize}
\item Max-battery: UE with the highest remaining battery in the potential helper list is selected. This algorithm minimizes the impact to the helper.
\item Proximity: UE closest to UE1 is selected. Proximity can be derived from the received signal strength of the discovery signal. This algorithm minimizes the energy consumption of the requester.
\item Virtual currency: To enforce fairness amongst all BDS participants, a virtual currency system can be set up where users pay for each help session. In this case, the helper selection algorithm can be thought of as a bidding system. This algorithm can be used to manage user incentive.
\end{itemize}
By the end of this helper selection process, a UE is chosen to help UE1 (UE3 in Figure~\ref{fig:bds_flow}). The AppSer sends this association to the D2D Server which implements the connection establishment procedure. In our terminology, UE1 is called the \emph{helpee}, and UE3 is called the \emph{helper}. Throughout the duration of the association, data from UE1 is relayed to the eNodeB through UE3.


\subsection{Overhead}
In our design, most of the energy consumed during the BDS initial association setup is carried by the core network. The participating UEs' energy expenditure is minimal. In addition to the control signaling associated with setting up the D2D connection, which will become standard once ProSe is put into practice, UE1 only spends energy on one request message (and an ACK) and a one-time discovery signal. Meanwhile, UE3 and other potential-helper UEs spend energy on a one-time report to the AppSer. This signaling energy cost is very small compared to a typical data session. In addition, the energy cost of these messages is much less than any distributed neighborhood discovery procedure (e.g. Wifi, bluetooth) in which multiple ``hello'' and ``reply'' messages have to be sent due to the lack of synchronization. Moreover, distributed optimal helper selection will require a lot more energy from the UEs than the centralized solution in BDS.

\subsection{Security implications}
Since BDS is a ProSe service, it has all security guarantees that will be offered by ProSe design. In particular, since encryption in LTE is done at UE1 and the eNodeB, UE3 sees only encrypted traffic. As a result, UE1's confidentiality is protected. Encryption also ensures that UE3 cannot insert its own messages into UE1's data stream. Thus data integrity is protected. A temporary ID (C-RNTI) is used instead of the real identity of UE1, therefore UE3 does not learn whose traffic it is carrying. Thus UE1's privacy is protected. The BDS does not incur any more security risk than what can already be obtained by an eavesdropper. 

User incentive is a more important consideration. The helper selection algorithm at the AppSer is responsible for ensuring fairness and discouraging selfish behavior. As mentioned above, one possible method is to use virtual currency to keep track of how much help a UE receives as well as how much help it has given. Designing such a mechanism is beyond the scope of this paper. Here we will assume that all UEs participate faithfully and show that cooperation indeed provides benefits.

\section{Performance evaluation framework}
\label{sec:performance_evaluation}
First, we want to reemphasize the role of network communications in smartphone battery consumption. To study the energy consumption breakdown on smartphones, researchers have either opened up phones and recorded power consumed by each component\cite{Carroll10}, or recorded total power consumption and switched on/off different components\cite{Huang12}. The consensus is that network communications and the display are the two biggest contributors, significantly higher than other components such as memory and processor. By reducing the power consumed by communications, our system provides a considerable gain for the overall battery life.

\subsection{Performance metric}
As described in the introduction, our motivation is to enable users to achieve higher utility for their smartphone's battery through cooperation. 
We define an expected usage time for a smartphone, and measure the probability that the user actually achieves that expected usage time in two cases: with and without cooperation. The probability that the user does not achieve his target usage time is called the \emph{probability of outage}. Since our system is a communication system, we only consider communication-related energy consumption.

In this work we consider a single macro-cell. The traffic model, mobility model, channel model and power consumption characteristics of the UEs are discussed next.

\subsection{Traffic model}
We use Poisson traffic models in our analysis. Poisson processes are very common in traffic modeling because they capture well the aggregate traffic caused by a large number of sources (in this case applications). Similar models were used by Nokia and Renesas Mobile Europe in their recent 3GPP contributions \cite{Nokia12, Renesas12}.  To determine the appropriate parameters for the Poisson processes, we consider a recent 3GPP technical report \cite{tr36.822}. In that report, 4 types of traffic scenarios are identified. Table~\ref{tab:traffic_scenarios} shows these traffic scenarios together with their uplink inter-arrival times and data sizes.
\begin{table}
	\centering
	\caption{Smartphone traffic scenarios}
	\begin{tabular}{|c|c|c|c|}
		\hline
		\multicolumn{2}{|c|}{\textbf{Traffic Scenario}} & \textbf{Inter-arrival time (sec)} & \textbf{Size (bytes)} \\ \hline
		\multicolumn{1}{|c|}{\multirow{2}{*}{Background}} & 
		\multicolumn{1}{|c|}{Light} & 10 & 50 \\ \cline{2-4}
		\multicolumn{1}{|c }{} &
		\multicolumn{1}{|c|}{Heavy} & 0.5 & 100 \\ \hline
		\multicolumn{2}{|c|}{Instant Messaging} & 2 & 100 \\ \hline
		\multicolumn{2}{|c|}{Gaming} & 0.1 & 25 \\ \hline
		\multicolumn{2}{|c|}{Interactive content pull} & 0.01 & 40 \\ \hline
	\end{tabular}
	\label{tab:traffic_scenarios}
\end{table}

We consider a smartphone usage scenario where the uplink data consists of 60\% light background, 20\% heavy background, 10\% instant messaging, 5\% gaming and 5\% interactive content pull traffic. We approximate these with an aggregate process. Similar to \cite{Nokia12, Renesas12}, we model the uplink data to arrive in bursts, with inter-arrival time equals 30 seconds. The size of each burst is modeled as a geometric random variable. The parameter for this geometric distribution is calculated so that the aggregate uplink data rate is equivalent to that of Table~\ref{tab:traffic_scenarios}.

\subsection{Power consumption}
In LTE, a UE's uplink transmit power (in dBm) is controlled by equation \eqref{eq:ul_pwr} (see \cite{Baker11_book_chap18, ts36.213}).
\begin{equation}
	P = \underbrace{P_0 + \alpha \PL}_{\text{open-loop}} + 
	      \underbrace{\Delta_{\text{TF}} + f(\Delta_{\text{TPC}})}_{\text{dynamic offset}} +
	      10\log_{10}(M)
	\label{eq:ul_pwr}
\end{equation}
The per-resource block (RB) power control consists of two components: a basic open-loop operating point and a dynamic offset. $M$ is the number of allocated RBs. 

$P_0$ is a semi-static nominal power level set by the eNodeB. $\alpha \PL$ is the path loss compensation component, where $\alpha$ controls the degree of compensation. $\PL$ is derived from the downlink Reference Signal Received Power. It includes shadowing but not fast fading. The dynamic control of UE uplink transmit power is designed to be an offset from the base operating point. This offset depends on two factors: the allowed modulation and coding scheme (TF stands for Transport Format) and a UE-specific transmitter power control (TPC) command.

In this work, we consider users to be homogeneous. Therefore we only use the open-loop power control in which $P_0$ is set to be the same for all UEs. Consequently, the transmit power of a UE depends only on its path loss (plus shadowing) and bandwidth.

In addition, after every data burst, the eNodeB lets the UE stay in RRC\_CONNECTED state for a little longer. In this state, the UE consumes notably more energy than RRC\_IDLE state. The duration that the UE stays in RRC\_CONNECTED state is configured by the eNodeB. We model this factor as well as other circuitry-related energy consumption as a constant component added to all transmissions (both D2D and regular uplink).

\subsection{Channel model}
We use WINNER II urban macro-cell model for our regular uplink connections, and WINNER II indoor model for our D2D connections \cite{winner}. Shadowing is modeled by lognormal distributions with parameters given in WINNER II documentation.

\subsection{Mobility}
In this work, we use a modification of the Random Waypoint Model to simulate user mobility. The Random Waypoint Model has a weakness that it favors the center of the cell more than the edge. Our modified model, which we call the Random Duration Model, generates a uniform distribution of user location. In this model, instead of choosing a new destination (waypoint) as a uniform random variable at each simulation step, a user chooses a random direction and a random travel duration, together with a random speed. We also implement a random pause time after each travel. This simulates the fact that in real life, people are not always moving. All of the mentioned distributions are uniform.

When a user reaches the cell boundary, we make that user reflect back into the cell (similar to how light reflects on a mirror). We choose this design to account for the fact that it is possible for D2D connections to exist between adjacent cells. It also allows us to use the same set of users throughout the simulation.

\section{Simulation results}
\label{sec:simulation}

\begin{figure*}
	\centering
	\begin{subfigure}[b]{0.45\textwidth}
                \includegraphics[width=\textwidth]{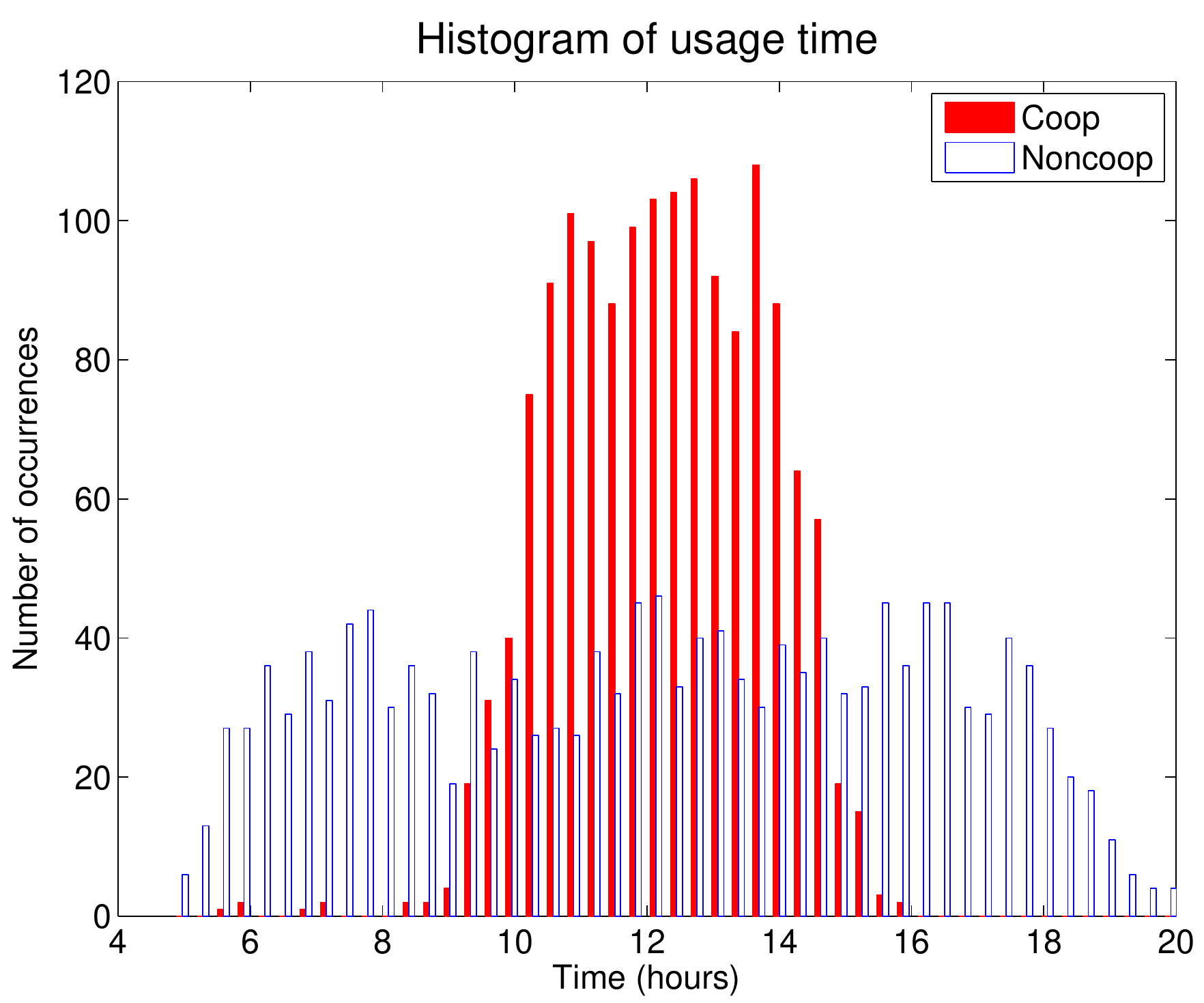}
                \caption{Histogram}
                \label{fig:pdf_usage_time}
        \end{subfigure}
        \quad 
        \begin{subfigure}[b]{0.45\textwidth}
                \includegraphics[width=\textwidth]{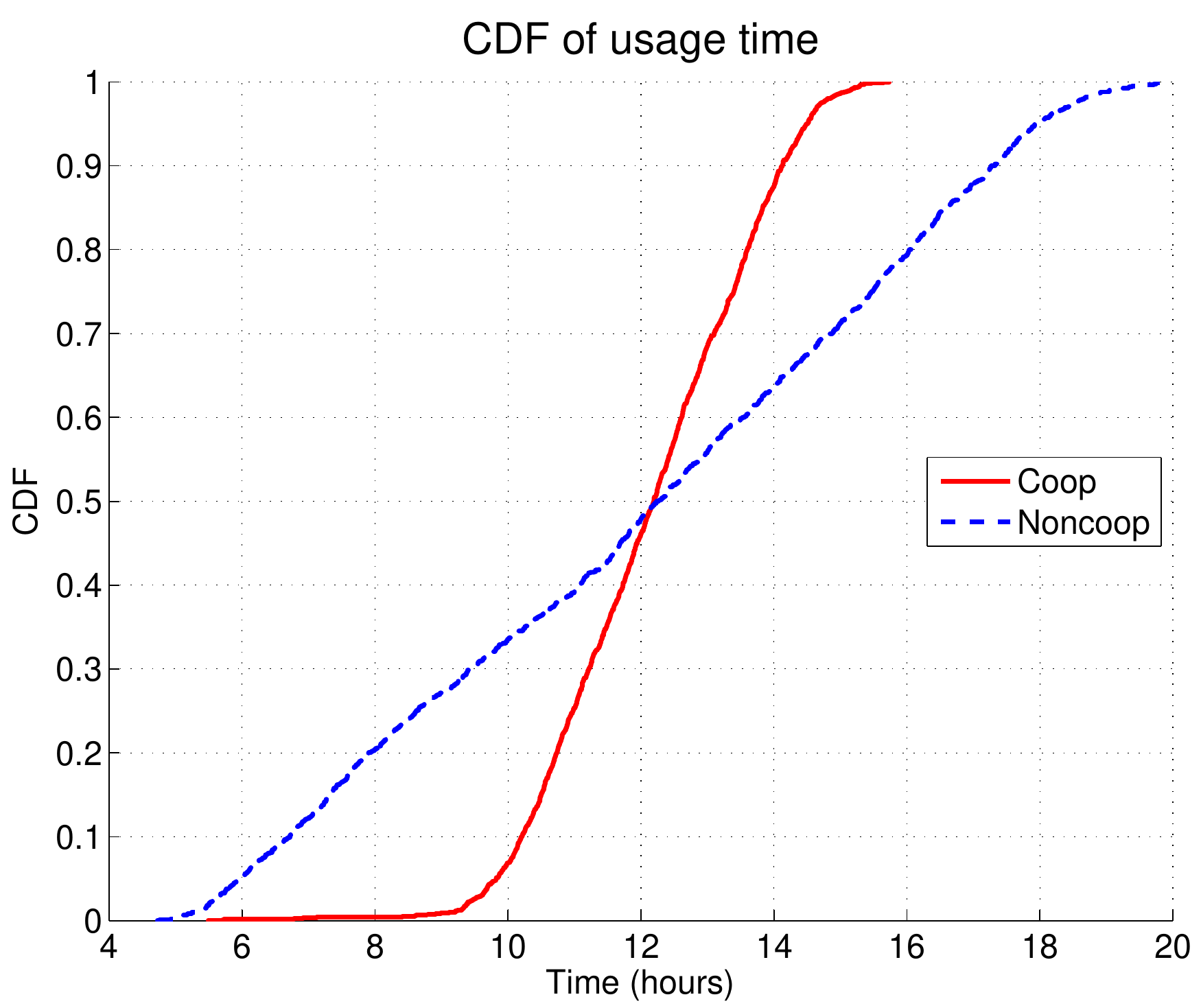}
                \caption{CDF}
                \label{fig:cdf_usage_time}
        \end{subfigure}%
	\caption{Distribution of usage time. Here the usage time of a UE is the period from the start of the simulation until that UE runs out of battery.}
	\label{fig:stats_usage_time}
\end{figure*}

We implemented an event-driven simulation in Matlab. The source code is available at \cite{github_bds}. The parameters that we use are summarized in Table~\ref{tab:sim_params}. The constant energy cost factor is derived from the report of power consumption in RRC\_CONNECTED state of UEs moving at 3 kmph with discontinuous reception period (DRX) set to 160 ms and release timer set to 5 seconds \cite{Nokia12}. The other parameters are also chosen to simulate a realistic scenario. 

The helper selection algorithm is based on proximity (the closest helper is selected). Besides $\gamma_1$ and $\gamma_2$, another factor influencing the degree of cooperation in the network is the signal strength of \texttt{BDSDiscovery}. This dictates the size of the neighborhood in which a user seeks help. We simulate the effect of this design parameter by fixing the radius over which a UE can find potential helpers. Our simulation is initialized with a snapshot of the network where the UEs are located at uniformly random locations within a circular cell. Each UE has a random battery level. We capture the instants that the UEs run out of battery for two cases, with and without cooperation. The statistics of time elapsed prior to exhaustion of the battery for the users are plotted in Figure~\ref{fig:stats_usage_time}. Figure~\ref{fig:pdf_usage_time} represents the distributions of the elapsed time for both cooperative and non-cooperative cases. In Figure~\ref{fig:cdf_usage_time}, we represent the CDF of these distributions.

\begin{table}[t]
	\centering
	\caption{Simulation parameters}
	\begin{tabular}{|c|c|}
		\hline
		\textbf{Parameter} & \textbf{Value} \\ \hline
		Cell radius & 500 m \\ \hline
		Number of UEs & 500 \\ \hline
		Mean data inter-arrival time & 30 s \\ \hline
		Mean burst size & 7800 bytes \\ \hline
		Speed & 0.1 - 3 m/s \\ \hline
		Pause duration & 0 - 300 s \\ \hline
		Walk duration & 30 - 300 s \\ \hline
		Path loss compensation factor $\alpha$ & 0.8 \\ \hline
		Constant energy cost factor & 15 mJ \\ \hline
		Communication battery budget & 300 J \\ \hline
		Base power $P_0$ & -69 dBm \\ \hline
		Maximum transmit power & 24 dBm \\ \hline
		Modulation order & QAM16 \\ \hline
		Code rate & 1/3 \\ \hline
		Carrier frequency & 2 GHz \\ \hline
		eNodeB antenna height & 25 m \\ \hline
		UE antenna height & 1.5 m \\ \hline
		Number of walls for indoor NLOS & 1 \\ \hline
		Cooperation threshold $\gamma_1,\gamma_2$ & 0.3, 0.3 \\ \hline
		Cooperation path loss threshold & 110 dB \\ \hline
		Cooperation radius & 30 m \\ \hline
	\end{tabular}
	\label{tab:sim_params}
\end{table}

The advantage of BDS can be clearly seen from Figure~\ref{fig:stats_usage_time}. As expected, our scheme redistributes the energy over the network, thus reducing the variation in the usage time. It can be seen that the mean lifetime of the battery for a user is roughly 12 hours for both the cooperative and non-cooperative cases. This is in accordance with the chosen simulation parameters. Let us consider a scenario wherein the users expect to use their phone for the normal work day, i.e. 10 hours. It can be seen that with our cooperative scheme, the probability of users not meeting their expectation (in other words, experiencing outage) is reduced from 35\% to 5\%. Similarly, if the target usage time is 8 hours, the probability of outage is reduced from 20\% to almost 0\%.

Figure~\ref{fig:cdf_usage_time} shows that if the intended usage is less than the mean usage duration provided by the battery, our scheme significantly improves user experience. However, for other cases our scheme is not beneficial. This is expected as the total amount of available energy in the network is constant. Our system makes the tradeoff of providing uniform utility to all users at the expense of shortening the usage time of some users with low traffic. However, as we have described, any remaining battery after the expected usage period is valueless. Thus even though the lifetime is shortened, it does not degrade the users' experience. 

We show the average amount of valueless battery for ``survived'' UEs as a percentage of the total battery capacity for various expected usage durations in Table~\ref{tab:valueless_battery}. It can be seen that our system utilizes effectively the otherwise useless remaining battery of the users. 

We would like to note that our scheme incurs an additional overhead due to the energy spent in communicating over the D2D links. However, this overhead can be considered as taken from the valueless battery pool thus it does not influence the benefits of our system.

\begin{table}
	\centering
	\caption{Average amount of valueless battery after various expected usage durations}
	\begin{tabular}{|c|c|c|}
		\hline
		\multirow{2}{*}{\textbf{Expected usage duration (h)}} & \multicolumn{2}{|c|}{\textbf{Valueless battery}} \\ \cline{2-3}
		& \textbf{Cooperative} & \textbf{Non-cooperative} \\ \hline
		6 & 31\% & 33\% \\ \hline
		8 & 20\% & 28\% \\ \hline
		10 & 12\% & 24\% \\ \hline
	\end{tabular}
	\label{tab:valueless_battery}
\end{table}

\section{Conclusions}
\label{sec:conclusion}
In this paper we have proposed a cooperative system, the Battery Deposit Service, as a new solution to prolong smartphones' battery life. We introduce the notions of \emph{valueless} and \emph{valued battery}, being the available battery on a user's phone when he does or does not have access to a charger, respectively. Our system allows users to expend their valueless battery to help conserve valued battery for others. Users who receive help (\emph{helpees}) utilize low-cost D2D links to tunnel their traffic to the neighboring helpers. The helpers relay those data over the more expensive cellular links. In effect, the helpers carry the burden of communication energy cost for the helpees. Variation in usage ensures that a user will play both roles of helper and helpee at some different times. We describe how our system can be implemented as a 3GPP proximity service. We confirm that BDS reduces the probability of users not meeting their usage expectation (\emph{probability of outage}) through a realistic simulation. We make the simulator source code available to interested parties who want to enhance our results.

\bibliographystyle{IEEEtran}
\bibliography{references}

\end{document}